\documentclass{aa501}

\usepackage{graphics}

\begin{document}

\font\small=cmr8           
\font\petit=cmcsc10        

\font\bbf=cmbx10 scaled\magstep1 
\font\bbbf=cmbx10 scaled\magstep2 
\font\bbbbf=cmbx10 scaled\magstep3 

\def\subti#1{\par\vskip0.8cm{\bf\noindent #1}\par\vskip0.4cm}
\def\ti#1{\par\vskip1.6cm{\bbf\noindent #1}\par\vskip0.8cm}
\def\bigti#1{\par\vfil\eject{\bbbf\noindent #1}\par\vskip1.6cm}

\def\cbigti#1{\par\vskip2.0cm{\bbbf\noindent\centerline {#1}\par\vskip0.5cm}}
\def\cti#1{\par\vskip1.0cm{\bbf\noindent\centerline {#1}\par\vskip0.4cm}}
\def\csubti#1{\par\vskip0.5cm{\bf\noindent\centerline {#1}\par\vskip0.3cm}}

\def\doublespace {\baselineskip 22pt}        

\def\eqd{\buildrel \rm d \over =}    
\def\p{\partial}           
\def\px{\partial _x}           
\def\py{\partial _y}           
\def\pz{\partial _z}           
\def\pt{\partial _t}           
\def\ssum{\textstyle\sum}
\def\arr{\rightarrow}
\def\id{\equiv}
\def\eqv{\leftrightarrow}
\def\fol{\rightarrow}
\let\prop=\sim
\def\gapprox{\;\rlap{\lower 2.5pt            
 \hbox{$\sim$}}\raise 1.5pt\hbox{$>$}\;}       
\def\lapprox{\;\rlap{\lower 2.5pt            
 \hbox{$\sim$}}\raise 1.5pt\hbox{$<$}\;} 

\def\ang{\,{\rm\AA}}
\def\cm{\,{\rm cm}}
\def\km{\,{\rm km}}
\def\kpc{\,{\rm kpc}}
\def\second{\,{\rm sec}}     
\def\erg{\,{\rm erg}}
\def\ev{\,{\rm e\kern-.1em V}}
\def\kev{\,{\rm ke\kern-.1em V}}
\def\k{\,{\rm K}}
\def\K{\,{\rm K}}
\def\gauss{\,{\rm gauss}}
\def\SFU{\,{\rm SFU}}

\def\R{\,{\rm I\kern-.15em R}}
\def\N{\,{\rm I\kern-.15em N}}
\def\N{\,{\rm /\kern-.15em R}}

\def\mhz{\,{\rm MHz}}

\def\n{\noindent}
\def\lead{\leaders\hbox to 10pt{\hfill.\hfill}\hfill}
\def\a{\"a}
\def\o{\"o}
\def\u{\"u}
\def\infinit{\infty}
\def\upr#1{\rm#1}

\def\dd{D^{\left(2\right)}}
\def\cc{C_d ^{\left(2\right)} \left(\epsilon\right)}

\newcount\glno
\def\no{\global\advance\glno by 1 \the\glno}

\newcount\secno
\def\newsec{\global\advance\secno by 1}
\def\sec{\the\secno}

\newcount\chapno
\def\newchap{\global\advance\chapno by 1}
\def\chap{\the\chapno}


\title{MHD consistent cellular automata (CA) models}
\subtitle{II. Applications to solar flares}

\author{H.\ Isliker \inst{1} \and A.\ Anastasiadis \inst{2} \and 
L.\ Vlahos \inst{1}}

\offprints{H.\ Isliker}

\institute{
Section of Astrophysics, Astronomy and Mechanics \\
Department of Physics, University of Thessaloniki \\
GR 54006 Thessaloniki, GREECE \\
isliker@helios.astro.auth.gr, vlahos@helios.astro.auth.gr 
\and
Institute for Space Applications and Remote Sensing \hfill\break
National Observatory of Athens  \hfill\break
GR 15236 Penteli, GREECE   \hfill\break
anastasi@space.noa.gr \hfill\break}

\date{Received ...; accepted ...}

\abstract{
In Isliker et al.\ (2000b), an extended cellular automaton (X-CA) model for 
solar flares was introduced.
In this model, the 
interpretation of the model's grid-variable is specified, and 
the magnetic field, the current, and an approximation to the electric field
are yielded,
all in a way that is consistent with Maxwell's and the MHD equations.
The model also reproduces the observed distributions of total energy, 
peak-flux, and durations.
Here, we reveal which relevant plasma physical processes are 
implemented by the X-CA model and in what form, and what global 
physical set-up is assumed by this model when it is in its natural state 
(self-organized criticality, SOC).
The basic results are:
(1) On large-scales, all variables show characteristic quasi-symmetries: the 
current has everywhere a preferential direction, the magnetic field exhibits 
a quasi-cylindrical symmetry.
(2) The global magnetic topology forms either 
(i) closed magnetic field lines around and along a more or less 
straight neutral line 
for the model in its standard form, or 
(ii) an arcade of field lines above the bottom plane 
and centered along a neutral line, 
if the model is slightly modified. 
(3) 
In case of the magnetic topology (ii), loading can be interpreted as if there 
were a plasma which 
flows predominantly upwards, whereas in case of the magnetic topology (i),
as if there were 
a plasma flow expanding from the neutral line.
(4) The small-scale physics in the bursting phase represent localized 
diffusive processes, which are triggered when a quantity which is an 
approximately linear 
function of the current exceeds a threshold. 
(5) The interplay of loading and bursting in the X-CA model can be 
interpreted 
as follows: the local diffusivity usually has a value
which is effectively zero, and it turns locally to an anomalous value
if the mentioned threshold is exceeded, whereby diffusion dominates the quiet 
evolution (loading), until the critical quantity falls below the threshold 
again.
(6) Flares (avalanches) are accompanied by the appearance of localized, 
intense electric fields.
A typical example of the spatio-temporal evolution of the electric 
field during 
a flare is presented.
(7) In a variant on the X-CA model, the magnitude of the current is 
used directly in the instability criterion, instead of the approximately
linear function of it.
First results indicate that the SOC state 
persists and is only slightly modified: distributions of the released energy
are still power-laws with slopes comparable to the ones of the 
non-modified X-CA model, and the large scale structures, 
a characteristic of the SOC state, remain unchanged. 
(8) The current-dissipation during flares is spatially 
fragmented into a large number of dissipative current-surfaces of varying 
sizes, which are spread over a considerably large volume,
and which do not exhibit any kind of simple spatial organization as a whole.
These current-surfaces do not grow in the course of time, they are very 
short-lived, but they multiply, giving rise to new dissipative current-surfaces
which are spread further around. They show thus a highly dynamic temporal 
evolution.
\hfil\break
{}\hskip1.5truecm It follows that the X-CA model represents an implementation of
the 
flare scenario of Parker (1993)
in a rather complete way, comprising aspects from small scale physics 
to the global physical set-up, making though some characteristic 
simplifications which are unavoidable in the frame-work of a CA.
}

\maketitle

\section{Introduction}

There are two approaches to modeling the dynamic evolution of solar flares: 
Magnetohydrodynamic (MHD) 
theory and Cellular Automaton (CA) models. MHD represents the traditional 
physical approach, being based on fluid theory and Maxwell's equations. It 
gives detailed insight into the small-scale processes in active regions, but
it faces problems to model the complexity of entire active regions and solar 
flares, so that it is usually applied to well-defined, simple topologies, or 
it is restricted to model only small parts of active regions, often in reduced
dimensions 
(see e.g.\ 
Mikic et al.\ 1989;
Strauss 1993;
Longcope \& Sudan 1994;
Einaudi et al.\ 1996;
Galsgaard \& Nordlund 1996;
Hendrix \& Van Hoven 1996;
Nordlund \& Galsgaard 1997;
Dmitruk \& Gomez 1998;
Galtier \& Pouquet 1998;
Georgoulis et al.\ 1998; 
Karpen et al.\ 1998;
Einaudi \& Velli 1999).
Global MHD models for solar flares are still in a rather qualitative state. 
CA models, on the other hand, 
can rapidly  
and efficiently treat complexity, i.e.\ spatially extended, large systems, 
which consist of 
many sub-systems (sub-processes), at the price, however, of 
simplifying strongly 
the local small-scale processes. Despite this, they are successful in 
explaining observed statistics of solar flares (the distributions of total 
energy, peak flux, and durations of observed hard X-ray time-series),
giving, however, no information or insight into the small-scale processes
(e.g.\ Lu \& Hamilton 1991, Lu et al.\ 1993, 
Vlahos et al.\ 1995, Georgoulis \& Vlahos 1996, Galsgaard 1996, Georgoulis \& 
Vlahos 1998; in the following, we will term these models or modifications of 
them {\it classical} CA models; a different category of models form the 
completely stochastic CA models for solar flares (e.g.\ MacPherson \& 
MacKinnon 1999), which we are not refering to in the following).

The classical CA models were originally derived in analogy to theoretical 
sand-pile models (Bak et al.\ 1987; 1988), and despite 
a vague association of the model's components with
physical variables and processes, they had to be considered as basically
phenomenological models. Later, Isliker et al.\ (1998) showed that 
the basic small-scale processes of the classical CA models can be 
interpreted as (simplified) MHD processes, for instance loading as strongly 
simplified shuffling, and
redistributing (bursting) as local diffusion processes. 
However, the classical CA models, even when interpreted in the way 
of Isliker et al.\ (1998), show still a number of unsatisfying points
from the point of view of MHD:
For instance, consistency 
with MHD and Maxwell's equations is unclear ($\nabla \vec B$ can not be 
controlled), secondary quantities such as currents and electric 
fields are not available.

In Isliker et al.\ (2000b; hereafter IAV2000), we introduced
the {\it extended} CA model (hereafter: X-CA model) for solar flares, 
in which the MHD-inconsistencies are removed, and which is 
more complete in the sense of MHD than the classical CA models.
The X-CA model consists in the combination of a classical CA model 
with a set-up which is super-imposed onto the classical CA, and which, 
concretely, yields the following benefits:
(i) The interpretation of the 
grid-variable is specified, turning the CA models therewith from 
phenomenological models 
into physically interpretable ones;
(ii) consistency with Maxwell's and the MHD equations is guaranteed,
and (iii) all the relevant MHD variables are yielded in a way consistent 
with MHD: the magnetic field 
(fulfilling $\nabla \vec B=0$), the current, and an approximation to the 
electric field. The 
set-up is super-imposable in the sense that it does not interfere with the 
dynamic evolution (the evolution rules) of the CA model it is super-imposed
onto, unless wished.  
The solar flare X-CA model 
is able to deal with the complexity of 
active regions, as are the classical CA models, but its components are now
physically interpretable in a consistent way. It represents a 
realization of plasma-physics (mainly MHD) in the frame of a CA model.

The X-CA model of IAV2000, which uses classical, existing models and 
extends them,
is to be contrasted to the construction of completely new CA models, derived
from MHD so that they are compatible with MHD (as for instance the recently
introduced CA model of Longcope and Noonan (2000),  
and the models of Einaudi \& Velli (1999),
and Isliker et al.\ (2000a), which 
moreover are of a non-SOC type).

In IAV2000, 
some basic properties of the X-CA model (in different variants) in 
its natural state (self-organized criticality, SOC) were revealed. In 
particular, it was shown that the observed distributions of total energy, 
peak-energy, and durations are as well reproduced by the X-CA model as 
they are by the classical CA models.
In this article, our aim is to reveal
the global physical set-up and the plasma-physical processes the X-CA 
model implements and represents when it is in the state of SOC.
These physical aspects of the X-CA model will be compared to the 
flare scenario suggested by Parker (e.g.\ Parker 1993; see also App.\ A). 
We will actually show that the X-CA model may be viewed as an 
implementation of Parker's (1993) flare scenario.

Differently, we may state the scope of this article as follows:
The X-CA model has at its heart a classical, 
phenomenological CA model, extends it yet and 
makes it physically interpretable. The X-CA model is thus a
physical CA model, contrary to the classical CA models.
It is now a posteriori to be seen what physical processes and structures
the X-CA actually represents.
It did, for instance,
not make sense (and actually was impossible) to ask for the magnetic topology 
implemented by the classical
CA models. Now questions like this one make sense, but the answers are not 
a priori given, and they are not contained in the frame of the classical
CA models alone. Also in this sense, the X-CA model represents a true 
extension of the classical CA models. Moreover, it is a priori not 
clear that the physical properties of the X-CA model we are going 
to reveal are compatible with what
is believed to happen physically in flares, just  
the statistical results are known to be compatible 
with the observations. 
The results 
of this article will yet show that the X-CA model can indeed be 
considered as making  
physically sense in the context of the flare modeling problem, 
it may be viewed as a reasonable {\it physical} model for flares,
all the more with the modifications we will introduce.

The questions concerning the implemented plasma-physical processes and global
physical set-up we address in this article are (Sec.\ 3): 
(1) what the magnetic topology in SOC state represents, 
(2) what the loading process actually simulates,
(3) what physical small-scale processes are implied 
by the model's energy release events, 
(4) how the electric field evolves in space and time during flares.
More-over, in Sec.\ 4, the X-CA model is modified to be closer to the 
flare scenario of Parker by using directly the current in the instability 
criterion. 
Lastly, it will be shown how the regions of current-dissipation, which appear
during flares, are organized in space and time (Sec.\ 5).
We will start 
by giving a short summary of the X-CA model (Sec.\ 2).

\section{Short summary of the extended CA (X-CA) model}

The extended CA (X-CA) model, whose detailed description is given in IAV2000, 
uses a 3--D cubic grid and the local vector-potential 
$\vec A_{ijk} = \vec A(\vec x_{ijk})$ at the grid-sites $\vec x_{ijk}$
as the primary grid-variable. In order to calculate derivatives of 
the vector-potential, the latter is made a continuous function in 
the entire modeled volume by interpolating it with 3--D cubic splines. 
In this way, the magnetic field is determined as
$\vec B = \nabla \wedge \vec A$, and the current as 
$\vec J = {c \over 4\pi} \, \nabla \wedge \vec B$, both as derivatives
of $\vec A$ and according to MHD.
The electric field is approximated by the resistive term of Ohm's law 
in its simple form, $\vec E = \eta \vec J$
(see the discussion of this approximation in Sec.\ 3.4),
where the diffusivity $\eta$ is given as $\eta=1$ at the bursting 
sites and zero everywhere else 
(following the analysis of Isliker et al.\ (1998); see also Sec.\ 3.3).

As a measure of the stress $\vec S_{ijk}$ in the primary field 
$\vec A_{ijk}$ we use two alternative definitions:  
(i) in Sec.\ 3 the classical or {\it standard} form  
$\vec S_{ijk} \equiv d\vec A_{ijk} := \vec A_{ijk}-{1\over n_n} \sum\limits_{n.n.} \vec A_{n.n.}$
(where the sum is over the first order nearest neighbours 
of the central point, and $n_n$ is the number of these neighbours),
following Lu \& Hamilton (1991) and most of the classical CA models; 
and (ii), in Secs.\ 4 and 5, taking advantage of the availability of secondary 
variables in the X-CA model, we use the current 
as a stress measure,
$\vec S_{ijk} \equiv \vec J_{ijk}$, which is physically more sensible 
than the standard $d\vec A_{ijk}$ (see the discussion in Sec. 4).

The grid-variable $\vec A$ undergoes two different regimes of dynamic 
evolution, loading (quiet evolution) and bursting (redistributing): 
During loading, random vector-field increments $\delta \vec A_{ijk}$
are dropped at random grid-sites. 
If locally the magnitude of the stress $\vec S_{ijk}$ 
exceeds a threshold then the system starts bursting: 
The vector-field is redistributed among the unstable site and its nearest
neighbours 
($\vec A_{ijk} \to \vec A_{ijk} - n_n/(n_n+1)\, \vec S_{ijk}$ 
for the central unstable grid-point, and 
$\vec A_{nn} \to \vec A_{nn} + 1/(n_n+1)\, \vec S_{ijk}$ 
for its nearest neighbours).
The amount of energy released in one burst is estimated as Ohmic
dissipation, $E_{burst} \sim \eta \, \vec J^2$  with, as stated, 
$\eta = 1$ at bursting sites 
(for details see Eq.\ (10) in IAV2000).

The model shows a transient phase before reaching 
a stationary state, the state of self-organized criticality (SOC), 
in which avalanches (flares) of all sizes occur, with 
power-law distributions of total energy, peak energy and durations,
which agree as well with the corresponding observed distributions as 
do the distributions yielded by the classical CA models (see IAV2000).

One of the necessary conditions for the system to reach the state of
SOC is that the loading increments 
$\delta \vec A_{ijk}$ exhibit a preferred spatial directionality 
(see e.g.\ Lu \& Hamilton (1991)). The used preferred direction
can be freely chosen, it does not change the statistical
results of the model. 
In Sec.\ 3, it will yet turn out that the used preferred direction
influences the magnetic topology.
We will investigate two preferred directions:
(a) parallel to the spatial diagonal of the simulation cube, 
as used in all
the classical CA models, and ultimately following the original prescription 
of Lu \& Hamilton (1991). We call this the {\it standard} 
preferred direction. (b) We will use the 
$x$-direction as preferred direction of loading.

The magnetic topology depends also on the boundary
conditions (b.c.) applied around the simulation cube;
actually it is the combination of the b.c.\ with 
the preferred direction of the loading increments which
determines the magnetic topology,
as will be shown in Sec.\ 3.
We will apply two different kinds of b.c.:
(1) open
b.c.\ (together with the standard preferred direction 
of the loading increments), as introduced by Lu \& Hamilton (1991) and used 
(most likely) in all the classical CA models, which we call thus the 
{\it standard} b.c. (2)
We will apply open b.c.\ around 
the simulation box except at the lower ($x$-$y$) boundary
plane, where we will assume closed b.c.\ 
(in combination with the preferred loading direction along the $x$-axis).
In App.\ B, the details of our implementation of open and closed b.c.\ 
are described.

\section{The physical processes and global physical set-up
         implemented by the extended CA model}

\subsection{The global topology of the magnetic field and of the current}

In IAV2000, it was demonstrated that the solar 
flare X-CA model exhibits a characteristic large scale organization of 
$\vert\vec B\vert$, the magnitude of the magnetic field, whereas the 
magnitude of the current, $\vert\vec J\vert$, seems not to exhibit any 
obvious large scale-organization. 
The question we address here is what these structures represent 
and whether they can be identified with structures in observed active regions.

The X-CA model makes magnetic field-lines available:
through the continuation (interpolation, see Sec.\ 2), the vector-potential is 
given also 
in-between grid-sites, hence also its derivatives, and therewith as well the 
magnetic field (see IAV2000 for more details). Magnetic field-lines at a 
fixed time can 
then be constructed as usual by integrating along the continuously given 
magnetic field, starting
from some initial point.

\subsubsection{The quasi-symmetries and their origin}

\begin{figure}
\resizebox{\hsize}{!}{\includegraphics{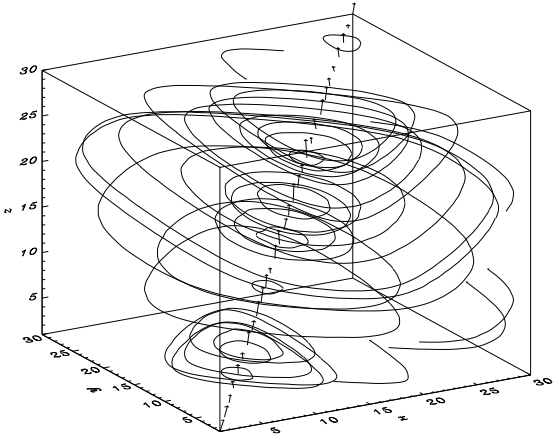}}
\caption{Magnetic field lines yielded by the X-CA model in its 
standard form, 
originating from randomly selected points. 
The vectors 
along the diagonal represent the (rescaled) currents 
(off-diagonal currents are
not shown). Near the diagonal a neutral line is situated.}
\label{}
\end{figure}

A typical single magnetic field line in the simulation box of the 
X-CA model in its standard form (see Sec.\ 2), 
which starts at an arbitrary point, 
winds itself around the diagonal and closes on itself, or it leaves the
modeled cube. In Fig.\ 1, a number of field lines is shown, starting 
from randomly chosen points in the simulation box (at an arbitrary, fixed 
time in the loading phase during the SOC state, and for a grid-size 
$30 \times 30 \times 30$). The magnetic field obviously shows cylindrical 
quasi-symmetry. 

Fig.\ 1 also shows the currents at the diagonal (the currents at the other 
grid-sites are not shown for purposes of better visualization): they are  
preferentially aligned with the diagonal, and this preferential direction 
is actually exhibited everywhere in the simulation box and at all times 
during SOC-state, so that also the current shows a quasi-symmetry. 

The reason for these quasi-symmetries is the quasi-symmetry 
imposed on the primary grid-variable by the loading rule:
The loading increments are asymmetric, namely with preferential direction 
parallel to the diagonal (Sec.\ 2).
Since the bursting rules are isotropic and symmetric in 
the three components of $\vec A$, the vector
potential $\vec A$ maintains the quasi-symmetry of the loading increments and
is preferentially aligned with the diagonal (parallel to 
$(1,1,1)$). As a result of this quasi-symmetry 
of the vector-potential, the magnetic field ($\prop \nabla\wedge\vec A$) and 
the current ($\prop \nabla\wedge\vec B$) must exhibit the mentioned symmetries:
If we introduce cylindrical coordinates, with the $z^\prime$-axis along the 
diagonal of the cube and $r$ the perpendicular distance from the 
$z^\prime$-axis, then, in obvious notation, due to its quasi-symmetry
$\vec A$ reduces to
$\vec A \approx A_{z^\prime}(r) \,\vec e_{z^\prime}$, from where it follows that $\vec B$
must be of the form
$\vec B = \nabla \wedge \vec A \approx  - {\partial A_{z^\prime} \over \partial r}\,\vec e_{\phi}$
(all the other terms vanish),
and finally for $\vec J$ we get
$\vec J = {c\over 4\pi} \nabla \wedge \vec B 
\approx  - {c\over 4\pi} {1\over r} {\partial\over\partial r} 
\left(r{\partial A_{z^\prime}\over\partial r}\right) \vec e_{z^\prime}$.

A consequence of these quasi-symmetries is that the current is always 
and everywhere
more or less perpendicular to the magnetic field, though in general
with a small parallel component, since the symmetries are always
slightly distorted.

\subsubsection{The magnetic field topology}

In the standard form of the X-CA model,
the magnetic field is obviously
described by quasi-cylindrically symmetric, closed field-lines around a 
more or less straight neutral line, which follows 
roughly the diagonal, as shown in Fig.~1. 

\begin{figure}
\resizebox{\hsize}{!}{\includegraphics{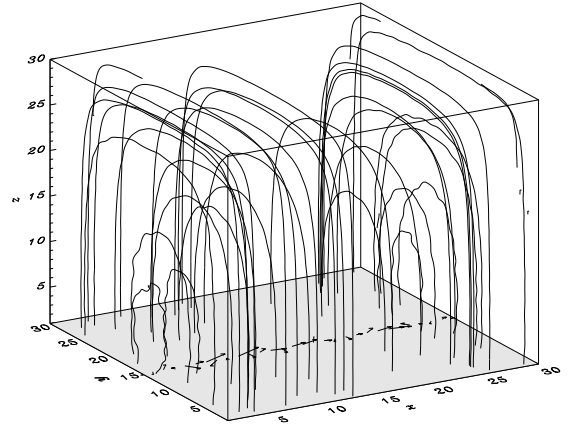}}
\caption{Magnetic field lines yielded by the modified X-CA model 
(see Sec.\ 3.1.2), 
originating from randomly selected points. The vectors 
shown in the shaded bottom plane represent the local 
(rescaled) currents (the currents at 
the other grid-sites are not shown). A neutral line is situated 
very roughly 
along the shown currents.}
\label{}
\end{figure}

A second, different magnetic topology is formed 
by the X-CA model in its non-standard form, where
we let the preferential direction of the loading increments 
be along the $x$-direction, and
we apply closed boundary conditions at the lower boundary (the 
$x$-$y$-plane),
keeping though all the other boundaries open (see Sec.\ 2).

The field lines form
now an arcade above the bottom (shaded) $x$-$y$-plane (Fig.\ 2), 
centered along a more or less straight 
neutral line in this plane (which follows very roughly the currents 
shown
in Fig.\ 2 --- note that, as in Fig.\ 1, only a subset of the currents
is shown, for better visualization). If we interpret the shaded 
$x$-$y$-plane as
the photosphere, then the picture is reminiscent of an arcade of loops.

The effect of the modifications on the magnetic
topology can be explained as follows:
The new preferred directionality of the loading increments causes the 
neutral line (the symmetry axis)
to be parallel to the $x$-axis, and to go through the 
mid-point of the grid 
(the argumentation is analogous to the one in Sec.\ 3.1.1).
The new boundary condition at the bottom plane causes the symmetry axis 
(neutral line) to move down into the 
bottom $x$-$y$-plane,
so that the field lines open and leave the simulation box through the 
bottom plane.

We just note that the statistical results the X-CA model yields in this
modification are still compatible with the observations 
(power-law distributions of peak-flux and total flux, with 
indices of roughly $1.8$ and $1.4$, respectively, i.e.\ the SOC state
persists).

\subsection{What the loading process simulates}

The interpretation of the loading process depends on 
the magnetic topology.
Let us 
first consider the variant of the X-CA model where 
the magnetic field forms 
an arcade of field 
lines, as in Fig.\ 2 (Sec.\ 3.1.2). 
The vector-potential $\vec A$ in coronal applications is in general assumed to 
evolve according to 
\begin{equation}
{\p \vec A \over \p t}
= \vec v \wedge \vec B + 
\eta {c^2\over 4\pi} \nabla^2\vec A - 
\eta {c^2\over 4\pi} \nabla(\nabla \vec A)
+\nabla \chi,
\end{equation}
which is the integrated induction equation of MHD, and where $\eta$ is the 
diffusivity and $\chi$ an arbitrary function. 
The loading process' role is to mimic the quiet evolution of active regions,
i.e., according to Parker's flare scenario, the shuffling of 
the magnetic field due to random foot-point motions (see App.\ A). In terms of
MHD, this implies that the convective term in Eq.\ (1) governs the temporal 
evolution.
Let us thus assume that the loading increments $\delta \vec A$ represent 
perturbations 
due to this convective term, 
i.e.\ $\delta \vec A \sim  (\vec v \wedge \vec B)$ (from Eq.\ 1), so that
the loading process implicitly implements the effect of a plasma with
velocity $\vec v$.
Since the increments of loading $\delta \vec A$  are preferentially along
the $x$-axis (see Sec.\ 3.1.2),
and since $\vec B$ is from left to right
in Fig.\ 2 --- note the preferential direction of the currents near the 
neutral line ---, the direction of $\vec v$ follows from the 
relation $\delta \vec A \sim  (\vec v \wedge \vec B)$ as being 
from the neutral line radially up- and outwards 
(radial in the sense of being perpendicular to the neutral line).
The sketch in Fig.\ 3 illustrates the situation.
Thus, the preferential direction of the loading can obviously be interpreted 
as if there were a plasma which flows
predominantly upwards, 
out of the shaded 
$x$-$y$-plane in Fig. 2 (see also Fig.\ 3).

\begin{figure}
\resizebox{\hsize}{!}{\includegraphics{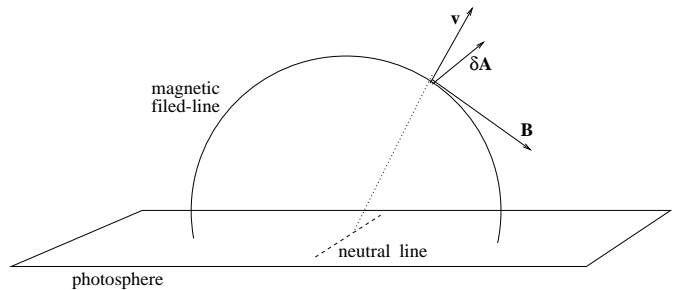}}
\caption{Sketch to illustrate the loading process:
the loading increments $\delta \vec A$ can be considered as being
proportional to $\vec v \wedge \vec B$, with $\vec v$ the velocity 
of the implicitely assumed plasma, and $\vec B$ the magnetic field.}
\label{}
\end{figure}

In case of the X-CA model in its standard form,
the magnetic topology 
(closed magnetic field lines around a straight neutral line, as in Fig.\ 1) 
would 
imply, by the same argumentation as before, that the loading must be 
considered as if there were a plasma expanding perpendicularly
away from the neutral line, symmetrically into all radial directions.

In conclusion,
the loading 
increments $\delta \vec A$ can be interpreted as being parallel to 
$\vec v \wedge \vec B$, with $\vec v$ the velocity of an assumed up- or 
out-flowing plasma, respectively, and, as a consequence, the direction of 
$\delta \vec A$ 
depends on the direction of
$\vec B$, the pre-existing magnetic field 
(not, however, on $\vert\vec B\vert$, the magnitude of $\vec B$).
--- Note that this interpretation is valid only in SOC state, when 
the magnetic field has organized itself into its characteristic large-scale
structure.

\subsection{Small scale processes: bursts}

Isliker et al. (1998) have shown that the redistribution (burst) rules 
we use (see Sec.\ 2)
can be interpreted as $\vec A$ evolving 
in the local 
neighbourhood of an unstable site according to the simple diffusion
equation
\begin{equation}
{\p \vec A \over \p t} =   \eta \, \nabla^2\vec A ,
\end{equation}
with the boundary condition
$(\vec n \nabla) \vec A = 0$ around the local neighbourhood,
and with diffusivity $\eta =1$. It is important to stress, however, that the 
X-CA redistribution 
rules for $\vec A$ 
do not represent the 
discretized version of Eq.\ (2), but they represent the transition {\it in 
one time step} from a given initial local field to the asymptotic solution 
of Eq.\ (2) (see Isliker et al.\ 1998). The time-step $\Delta t$ of the 
X-CA model therewith is roughly the diffusive time, and the grid-spacing 
$\Delta h$ is roughly the diffusive length scale
(as the value of the diffusivity, the numerical values of $\Delta t$ and 
$\Delta h$ are not specified and set to one).

\begin{figure*}
\resizebox{\hsize}{!}{\includegraphics{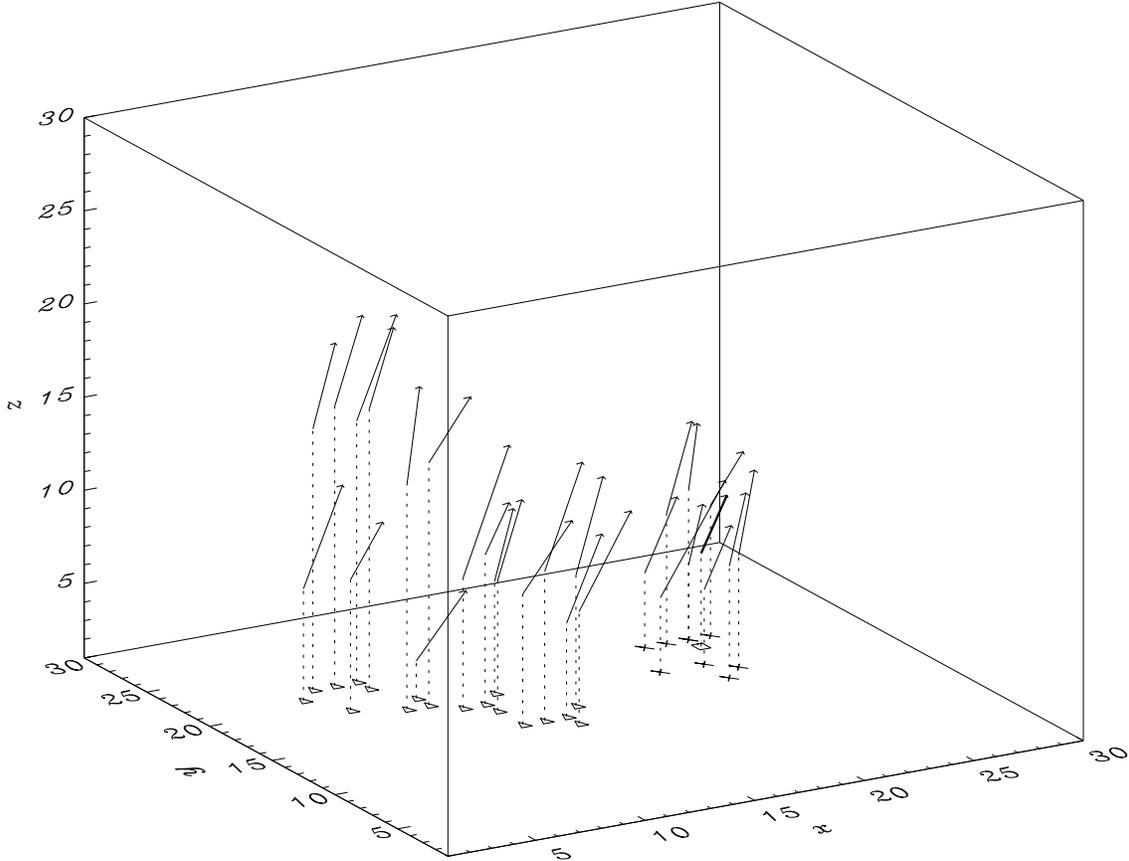}}
\caption{The electric field-vectors during a flare, at three different 
time-steps: at the beginning of the flare (bold-vector, projected
grid-site in $x$-$y$--plane marked with a rectangle); after nine time-step
(marked with 'x'); after 91 time-steps (marked with triangles). The vectors 
are shown in 3--D parallel projection, rescaled for visualization purposes, 
with length proportional to $\vert \vec E \vert$. Note that the electric 
fields of three different time-steps are shown together for visualization
purposes, in the model actually only one set appears at a time, the fields of 
the previous time-steps have become zero again, at later times.}
\label{}
\end{figure*}

The evolution of $\vec A$ according to Eq.\ (2) in the X-CA corresponds 
exactly to what the induction equation of MHD (Eq.\ 1) is expected 
to reduce to
for the case of anomalous
diffusion in cylindrical symmetry:
(a) According to Parker's flare scenario, the diffusivity 
at unstable sites is anomalous, i.e.\ increased by several orders of 
magnitude (see App.\ A), so that the convective term can be assumed to 
be negligible in the induction equation. 
(b) The quasi-symmetry of the vector-potential (Sec.\ 3.1.1) implies that
$\vec A$ is of the form $\vec A \approx A_{z^\prime}(r) \vec e_{z^\prime}$ 
(by using the same cylindrical coordinate system as in Sec.\ 3.1.1),  
so that 
$\nabla \vec A \equiv (1/r) \,\p/\p r \,(rA_r) + (1/r)\,\p A_\phi/\p \phi + 
\p A_{z^\prime}/\p z^\prime \approx 0$, and therewith 
$\nabla(\nabla \vec A) \approx 0$ in the induction equation.

The most characteristic simplifications made by the 
X-CA model are:
(i) The boundary conditions are unrealistically simple. They actually imply 
that $\int_{n.n.}\vec A\,dV$ is conserved in the diffusion events
(see Isliker et al.\ 1998).
(ii) All the diffusion events have the same diffusivity, diffusive length 
scale and diffusive time.

The amount of energy
released in the diffusion events of the X-CA model 
is determined through the expression for Ohmic dissipation (see Sec.\ 2), 
following directly the MHD prescription.

Lastly, we turn to the instability criterion of the X-CA model in its 
standard form (the non-standard instability criterion is discussed 
in  Sec.\ 4):
Bursts occur in the model if the local stress 
($\vert d\vec A_{ijk}\vert$) exceeds 
a threshold (see Sec.\ 2)). In IAV2000, it has been shown that 
there, where the stress $\vert d\vec A_{ijk}\vert$ exceeds the threshold, also
$\vert \vec J_{ijk}\vert$ is increased, and after a burst both 
$\vert d\vec A_{ijk}\vert$ and $\vert \vec J_{ijk}\vert$ are relaxed. 
Actually,  
$\vert\vec J_{ijk}\vert$ is an approximately linear function 
of $\vert d\vec A_{ijk}\vert$
for large enough $\vert d\vec A_{ijk}\vert$, 
monotonically increasing with $\vert d\vec A_{ijk} \vert$ (see IAV2000). 
This is very reminiscent of Parker's flare 
scenario
(see App.\ A):
During the 
loading phase, a diffusivity $\eta =0$ is assumed everywhere. 
If a threshold in the stress, 
which is 
a linear function of the current, is reached somewhere, then $\eta = 1$ in the
local neighbourhood, 
and 
diffusion sets on. 
As in Parker's flare scenario, the diffusivity 
thus assumes anomalous values (one), if a linear function of the current 
reaches a certain threshold. 
Otherwise it is small (ordinary) and effectively set to zero.

\subsection{The electric field}

\begin{figure}
\resizebox{\hsize}{!}{\includegraphics{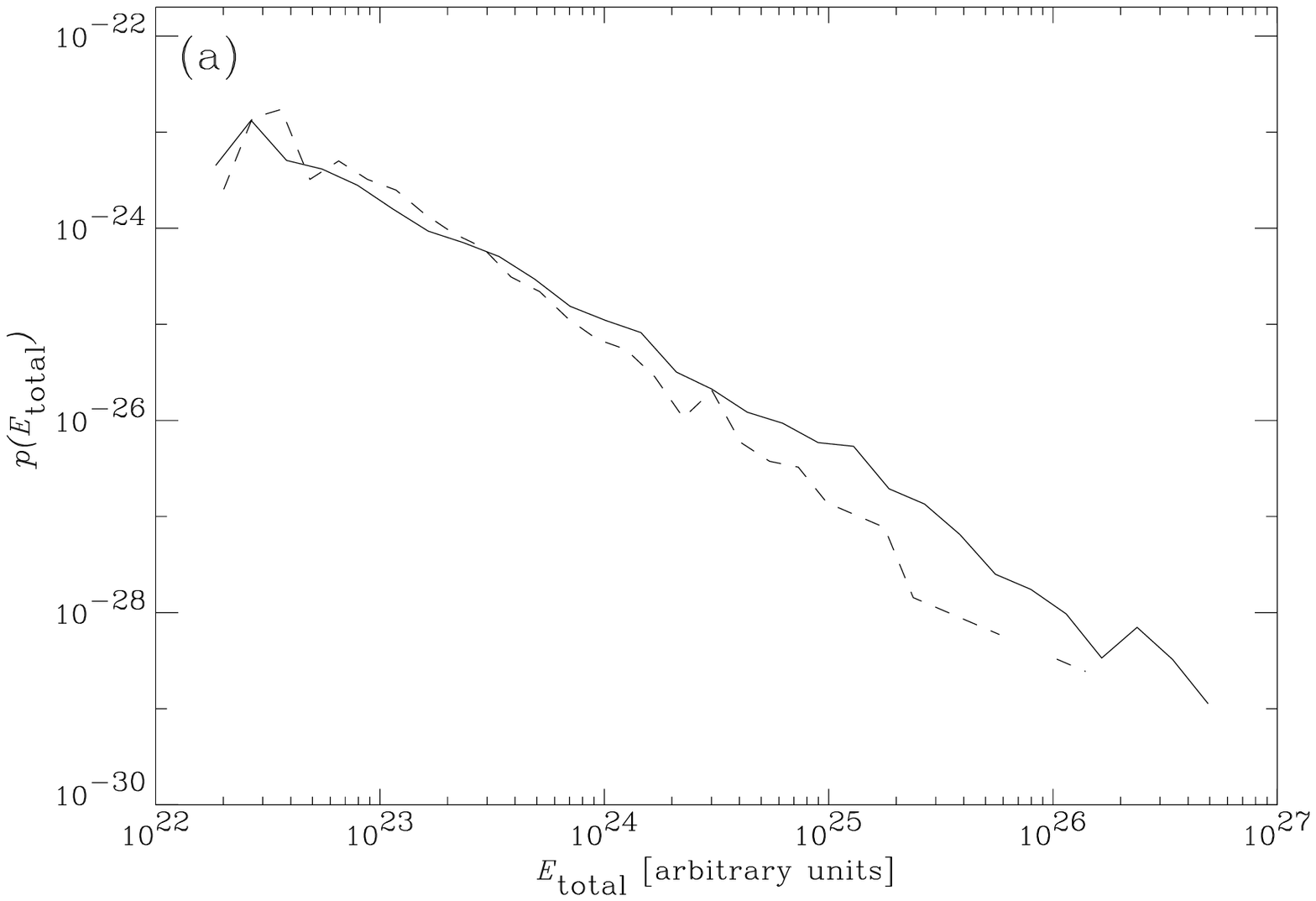}}
\resizebox{\hsize}{!}{\includegraphics{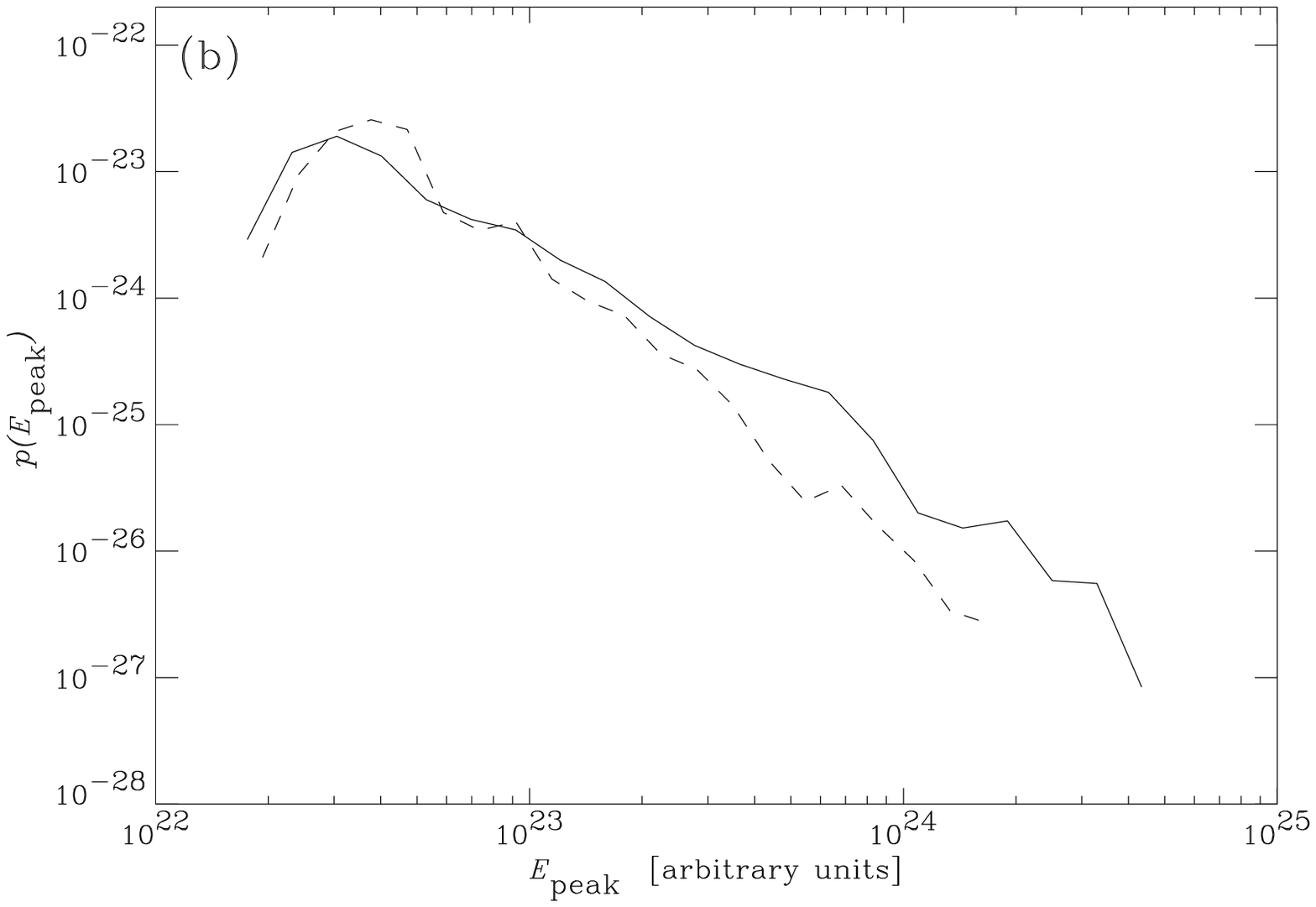}}
\caption{Probability distribution of total energy (a) and peak flux (b)
for the X-CA model in its standard form according to Sec.\ 2 (solid), 
and using the current in the instability criterion and 
in the redistribution rules, see Sec.\ 4 (dashed). 
The energy units are arbitrary.}
\label{}
\end{figure}

Of particular interest is the electric field in the X-CA model, 
since it is the cause for particle acceleration and the associated non-thermal 
radiation of flares. 
In the X-CA model,
the electric field is approximated by the resistive 
term of Ohm's law in its simple form, $\vec E=\eta\vec J$ (Sec.\ 2), 
which can be expected to be a good approximation,
since in 
the applications we are interested in events of current dissipation. 
This argument is actually based on Parker's 
flare scenario (see App.\ A), 
together with the assumption that Ohm's law in its 
simple form is a reasonable approximation in coronal active regions: 
the diffusivity is small at most times in active 
regions (build-up phase, loading phase), and the simple Ohm's law for the 
electric field 
($\vec E = \eta \vec J - {1\over c} \vec v \wedge \vec B$)
reduces to 
$\vec E = - {1\over c} \vec v \wedge \vec B$. 
However, if the diffusivity becomes anomalous at a bursting site, 
as described in App.\ A, and increases by several orders of magnitude, 
then the electric field must be expected to be dominated by the resistive 
part, $\vec E = \eta \vec J$,
and it is this contribution to the electric field which will be the cause of 
particle acceleration during flares.
We thus assume in our applications the $\vec E$-field usually to be zero
(assuming in the non bursting phase the velocities to be small and 
therewith the electric field to be negligible), and only
if the instability criterion is fulfilled at some grid-sites,  
an electric field of the form $\vec E=\eta\vec J$  appears for one time-step.
If the burst is over (in the following time-step, and if the site does not
again fulfill the instability criterion), the electric field is zero 
again.

In Fig.\ 4, the electric field as it appears during a flare
(avalanche) in the SOC state of the X-CA model
is illustrated (for a $30\times 30\times 30$-grid):
We chose a medium-size flare, which lasted 181 time-steps.
In the figure, the electric field is shown for three different time steps in 
the course of the flare: At the onset of the flare, one grid-site is unstable,
and it carries an electric field, 
whereas all the other grid-sites have a zero electric 
field. After nine time-steps, the instabilities have traveled away from
the initially unstable site and are spread around it, and the electric 
field appears correspondingly at these sites. 
After 91 time-steps, the unstable sites are spread over a larger volume,
which is not surrounding the initial site anymore, the instabilities have 
traveled to a different region in the grid, where the corresponding 
electric fields appear.

Remarkably, the electric-fields which appear are all of comparable 
intensity, 
and they are all more or less along the same preferential direction. 
The 
former is due to the fact that the current is an approximately linear 
function 
of $d\vec A$ for large values of $d\vec A$, as stated earlier 
(see IAV2000 for 
details), which itself is 
just above the threshold, so that through the relation
$\vec E=\eta\vec J$ all the electric field 
magnitudes are similar. The parallelity is due to the quasi-symmetry 
obeyed by the current in the SOC state
(Sec.\ 3.1.1):
the current is preferentially along the diagonal of the cubic grid, and as a 
consequence of the relation $\vec E=\eta\vec J$, 
the electric field has the same preferential
direction.

Likewise, the electric field is
always more or less perpendicular
to the magnetic field, exhibiting though in general a small parallel 
component. This is a again a consequence of the relation
$\vec E=\eta\vec J$ and of the corresponding
property of the current (see Sec.\ 3.1.1).

\section{A modification of the extended CA model: the current as the critical 
quantity}

One difference between the X-CA model in its standard form 
and Parker's flare scenario 
is that the current 
$\vert\vec J_{ijk}\vert$ is not directly used as a critical quantity 
(see App.\ A), but rather 
$\vert d\vec A_{ijk}\vert$ 
(see Sec.\ 2 and the discussion in Sec.\ 3.3). 
This leads us to modify the X-CA model, 
and to use as the stress measure $\vec S$ directly the current $\vec J$ 
(see Sec.\ 2).
The new instability criterion is  
\begin{equation}
\vert \vec J_{ijk} \vert > J_{cr} .
\end{equation}
(with $J_{cr} = f{c\over 4\pi}A_{cr}$, where $f$ is chosen from Fig.\ 4 of 
IAV2000 such that the threshold $A_{cr}$ for 
$\vert\vec J_{ijk}\vert$ corresponds 
roughly to the threshold for $\vert d\vec A_{ijk}\vert$).
Redistribution events in this variant can thus directly be considered as 
representing current driven instabilities. 
The use of $\vec J_{ijk}$ instead of $d\vec A_{ijk}$ also
in the redistribution rules is motivated through the following
argument: 
the use of $d\vec A_{ijk}$ can be justified by Eq.\ (2), which is
hidden behind the bursts in the X-CA model, since $d\vec A_{ijk}$ is an
approximation to $\nabla^2 \vec A$ (see IAV2000). 
However, since the induction equation (Eq.\ 1),
when neglecting the convective term, can equivalently be written 
as 
\begin{equation}
{\p \vec A \over \p t}
= -\eta c \vec J  + \nabla \chi ,
\end{equation}
it is more natural from the point of view of MHD to use $\vec J_{ijk}$
also in the redistribution rules.
The result of these modifications is (using a grid-size 
$30\times 30\times 30$) that according to first
results the SOC state persists, with power-law distributions (Fig.\ 5) which 
are a bit steeper (5\% to 10\%), 
and a large scale structure of the magnetic field which is very close 
to the one of the non-modified X-CA model (see Sec.\ 3.1, and IAV2000).

We just note that when using $\vec J_{ijk}$ only in the instability
criterion, but not in the redistribution rules (where still 
$d\vec A_{ijk}$ is used), it turns out that
sooner or later the model finds itself in an infinite loop, 
independent of the value of $J_{cr}$. 
The reason is that $\vert \vec J_{ijk}\vert$ is an approximately linear 
function
of $\vert d\vec A_{ijk}\vert$ only for large stresses 
$\vert d\vec A_{ijk} \vert$, 
but
the opposite is not true, there are cases where 
$\vert \vec J_{ijk} \vert$ is large but 
$\vert d\vec A_{ijk} \vert$ is almost zero (see IAV2000 for details). 
In these cases,
a burst should happen ($\vert\vec J_{ijk}\vert$ is large), 
but the almost zero $d\vec A_{ijk}$ 
cannot redistribute the fields, and the algorithm falls into an endless 
loop.

\section{The spatial organization of the current-dissipation regions}

Before turning to flares, it is worthwhile to illustrate how the spatial 
regions of intense, but sub-critical current are spatially organized during 
the quiet evolution 
(loading) of the X-CA model, since any structures the current forms in the 
quiet evolution are the base on top of which the flares take place.
A three-dimensional representation of the surfaces of constant current-density
at a sub-critical level
($\vert \vec J\vert = const. =9.1\,10^{10}$) is shown in Fig.\ 6, 
for an arbitrary time during the loading phase in the SOC-state (i.e.\ 
no grid-sites are unstable in the figure), as given by the X-CA model
in the version of Sec.\ 4. The current in the entire
simulation box ranges from $0.1.\,10^{10}$ to $12.0\,10^{10}$, and 
the threshold is $J_{cr} = 12.02\,10^{10}$ (the units are arbitrary).
The current-density obviously organizes itself into a large
number of current surfaces of varying sizes, all smaller though
than the modeled volume, and homogeneously distributed over the
simulation box.
The numerical values of the current densities span a range until just very 
little below the threshold, which is actually typical for the loading 
phase, and consequently the system can easily become unstable at some 
grid-site through further loading.

\begin{figure}
\resizebox{\hsize}{!}{\includegraphics{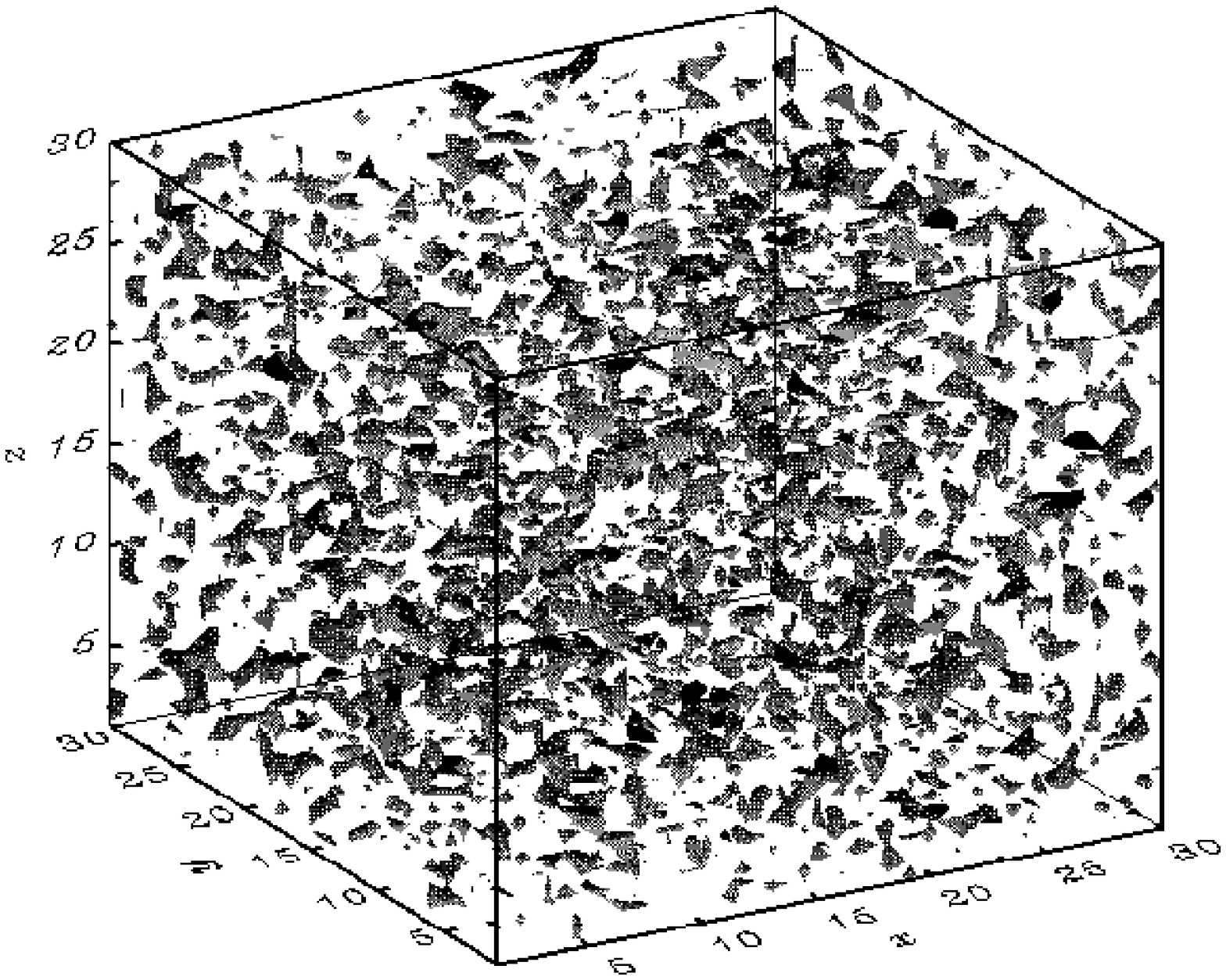}}
\resizebox{\hsize}{!}{\includegraphics{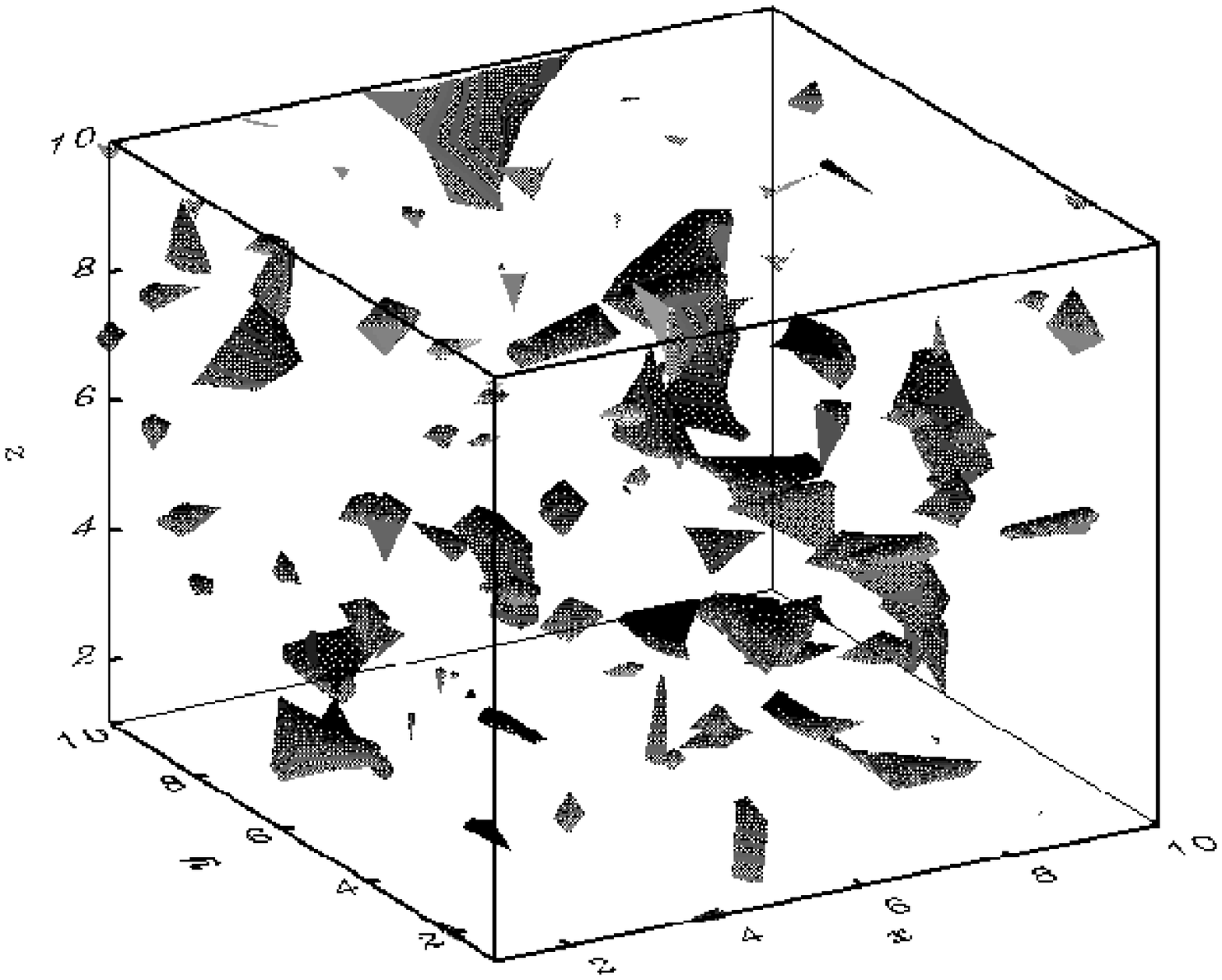}}
\caption{Three-dimensional representation of the (shaded) surfaces of
constant (sub-critical) current-density ($\vert \vec J\vert = 9.1\,10^{10}$) 
at an arbitrary time
during the loading phase, in the entire simulation box (top panel), and zoomed 
(bottom panel).
}
\label{}
\end{figure}

Of particular interest is the spatial structure of the
unstable regions during flares, i.e.\ of the regions of current-dissipation
(see Sec.\ 3.3), 
whether and how these regions are spatially
organized, and also how one spatial structure emerges from the immediately 
previous one.  
In Fig.\ 7, the regions of current-dissipation are
shown for two different time-steps during a flare 
(i.e.\ the surfaces of $\vert \vec J\vert =J_{cr} \equiv  12.02\,10^{10}$,
which enclose the regions where the current is above the threshold):
A flare starts with one single, usually very
small, region of super-critical current. This small region does not grow,
but multiplies in its neighbourhood, it gives rise to 
spreading of unstable regions, i.e.\ of current-dissipation regions. 
The secondary regions of current-dissipation multiply again, etc., and after
not too many time-steps the appearing current-dissipation regions become 
numerous and 
vary in size, the larger ones having the shape of
current surfaces, as in Fig.\ 7 (top panels), which is at 
an early stage in the flare.
These current-surfaces multiply further and travel through
the grid, giving rise now to even larger numbers of current
surfaces, as in Fig.\ 7 (bottom panels), which is 
at a later time, during the main phase of the flare. The degree of 
fragmentation has increased, 
and the current surfaces are spread now over a considerable volume.
The picture in Fig.\ 7 (bottom panels) is typical for a flare of intermediate
duration (the flare lasted 177 time-steps) as far as the size of the 
largest current surfaces, the degree of fragmentation, and the spatial
dispersion are concerned, though the concrete picture continuously changes 
in the course of time. 
Towards the end of the flare, the current
surfaces tend to become less numerous, and finally they die out quickly.

\begin{figure*}
\resizebox{\hsize}{!}{\includegraphics{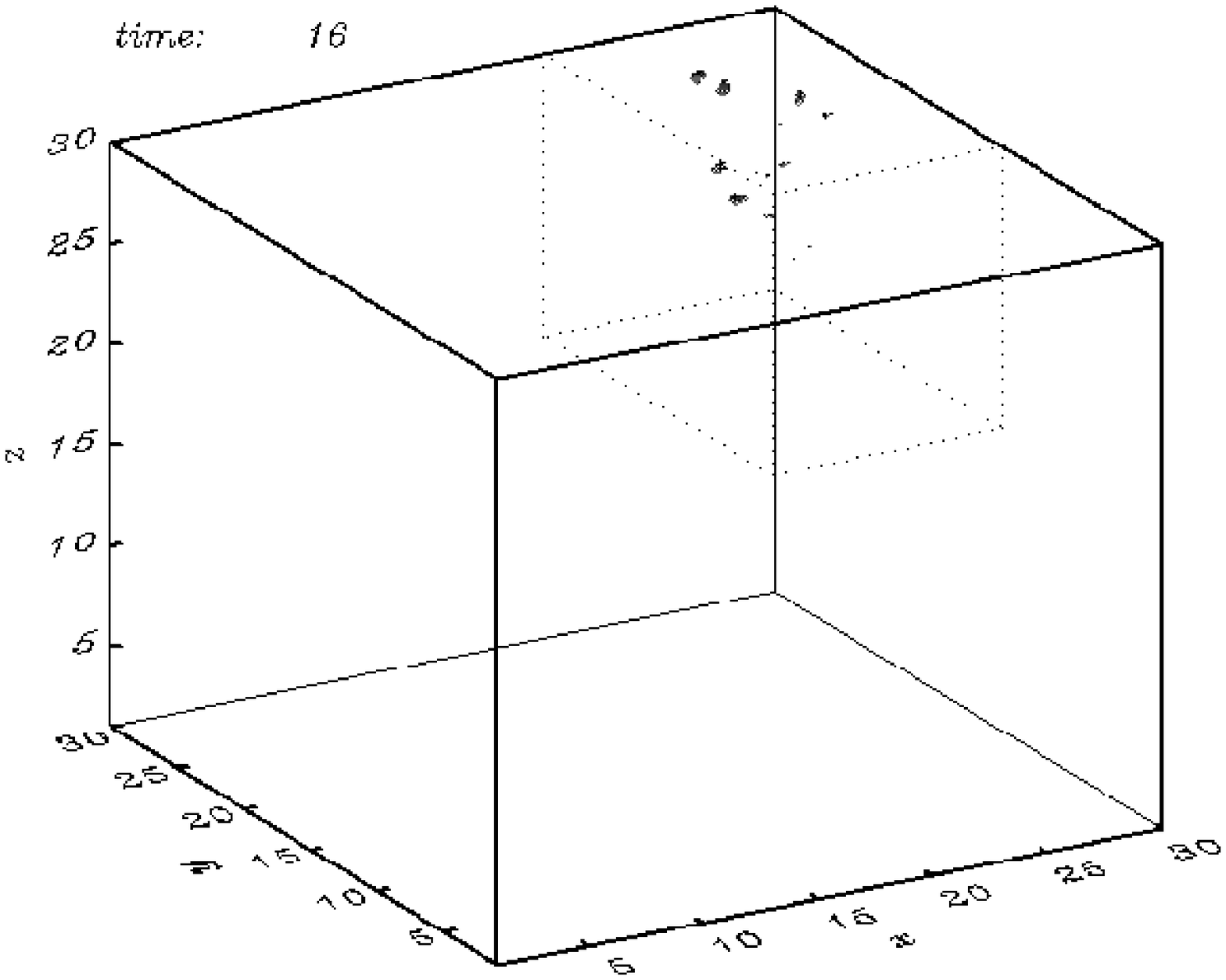}
                   \includegraphics{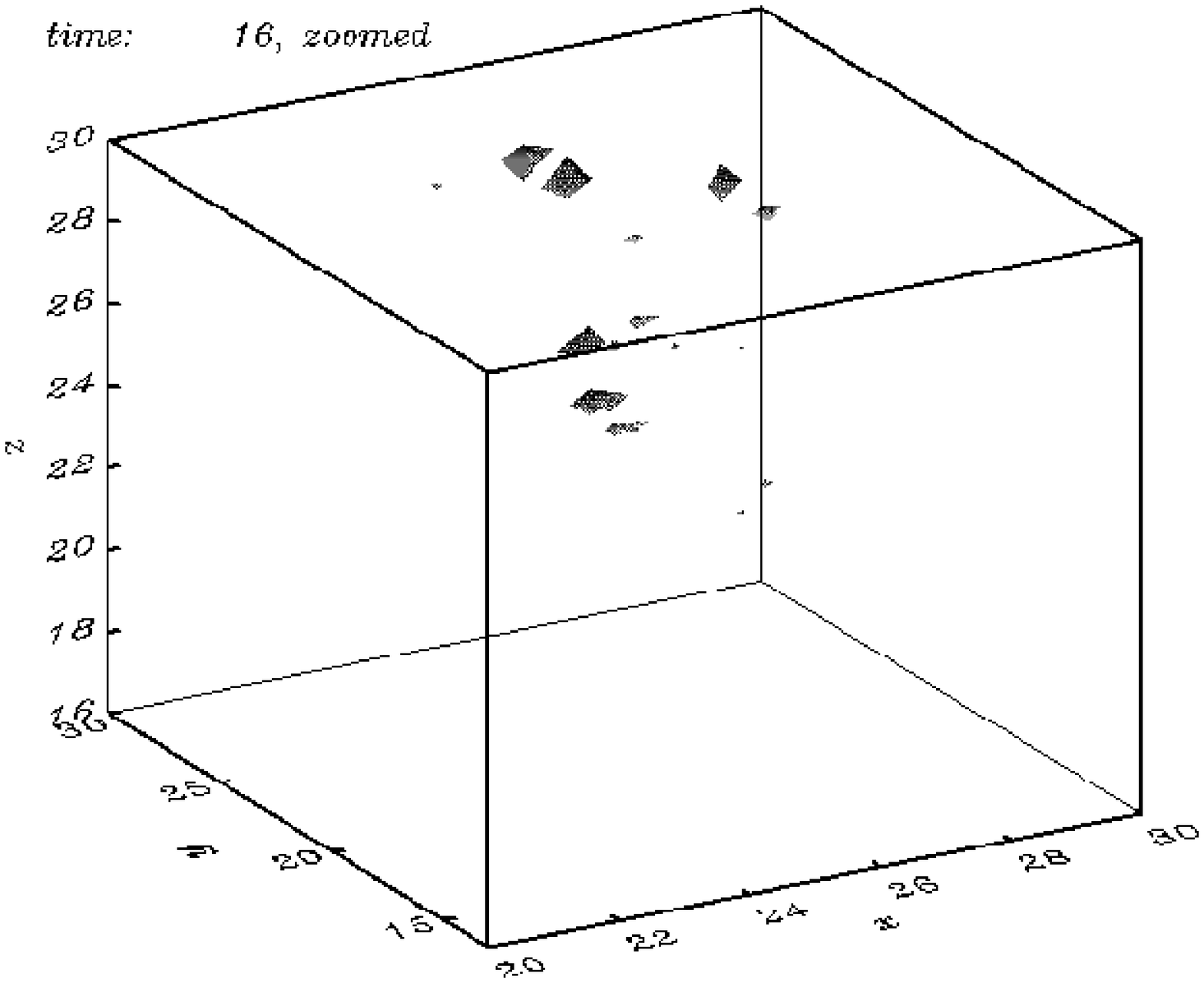}}
\resizebox{\hsize}{!}{\includegraphics{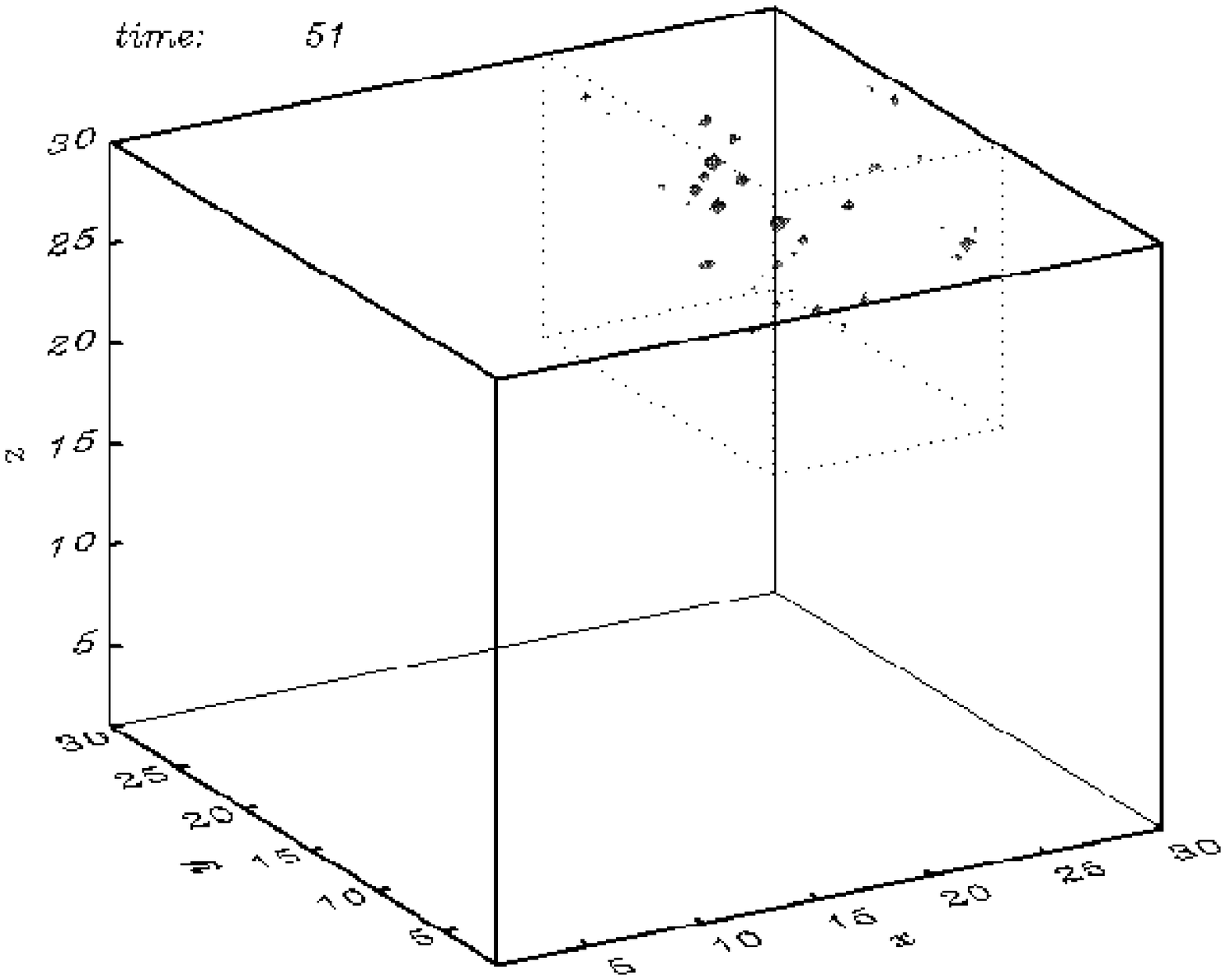}
                   \includegraphics{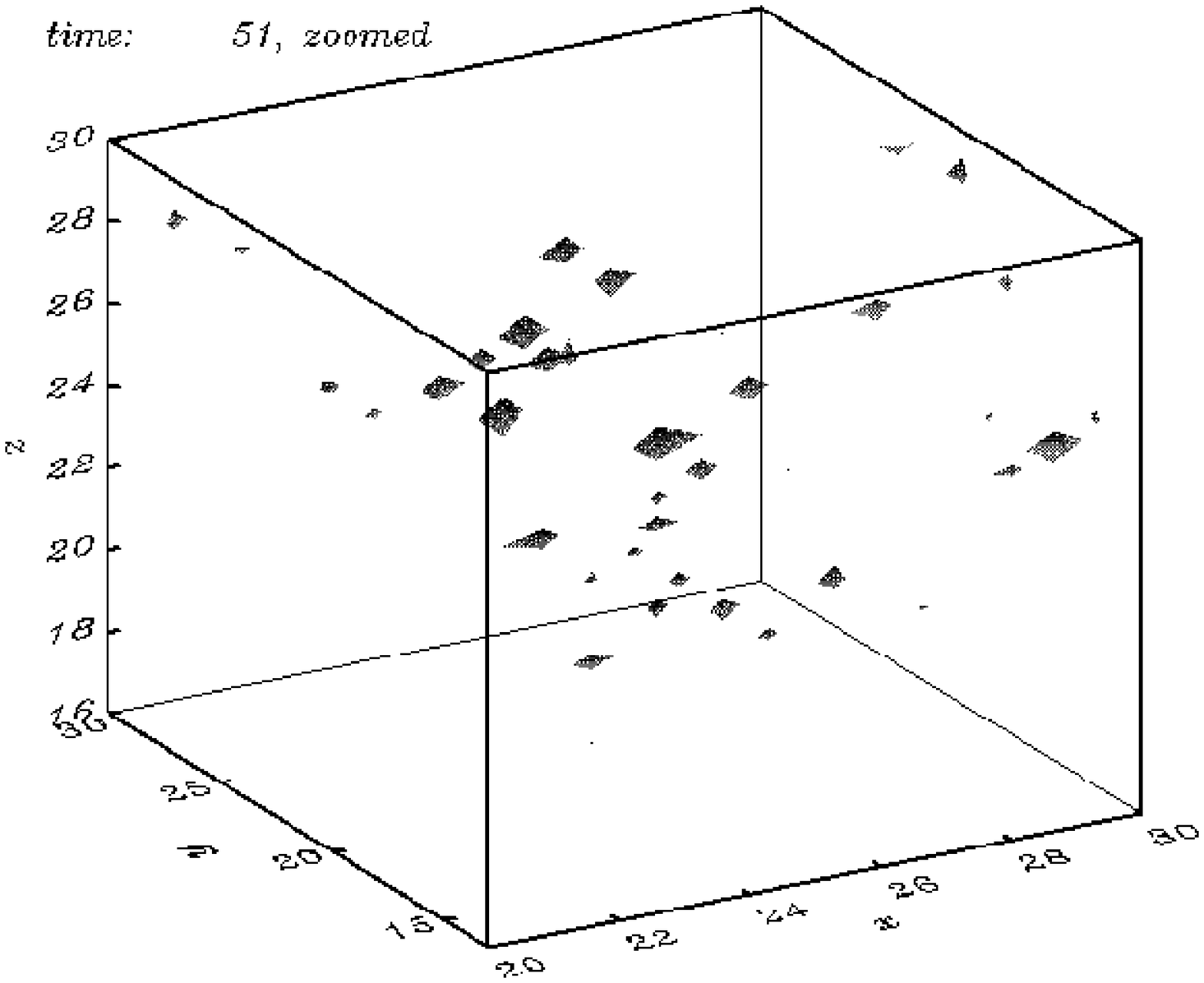}}
\caption{
Three-dimensional representation of the current-dissipation regions
appearing during a flare, i.e.\ of the (shaded) surfaces of
constant current-density equal to the threshold 
($\vert \vec J\vert = J_{cr} = 12.02\,10^{10}$),
for different times during a flare: at time-step 16 after the beginning
(top-panels, left: the entire simulation box, right: zoom of the dotted 
region), and
at time-step 51 (bottom-panels, left: the entire simulation box, 
right: zoom of the dotted region). 
}
\label{}
\end{figure*}

\section{Summary, Discussion and Conclusions}

\subsection{Summary}

The extended CA (X-CA) model, introduced in IAV2000, is consistent with 
Maxwell's
and the 
magnetic part of the
MHD equations, and makes all the secondary variables (currents,
electric fields) available. 
In IAV2000, it was shown that the X-CA model (in different variants) 
reproduces as well as the classical CA models the observed distributions
of total flux, peak-flux, and durations,
and that it can be considered as a model for energy release through
current-dissipation,
which was confirmed here and supported with more facts.
In this article, our aim was
to reveal 
the small-scale physics and the global physical set-up 
implemented by the X-CA model when it is in the SOC state. 
The basic results are:

\smallskip
\n {\bf 1.\ Quasi-symmetries of all the grid variables:} 
A consequence of the SOC state are the characteristic quasi-symmetries of the
fields: preferential alignment with the cube-diagonal for 
the vector potential and the current, and cylindrical quasi-symmetry around 
the diagonal for the magnetic field (for the model in its standard form).

\smallskip
\n {\bf 2.\ Magnetic field topology:}
For the preferred directionalities of loading and boundary conditions adopted here,
the global topology of the magnetic field has two varieties: 
either it forms  an arcade of magnetic field lines, 
centered along a neutral line for the modified X-CA model, 
or it forms  
closed magnetic field lines around and along a more or less straight 
neutral line for the model in its standard form.

\smallskip
\n {\bf 3.\ Interpretation of the loading process:} 
In the variant of the model where the magnetic field 
forms an arcade of 
field lines above a bounding surface which includes a neutral line, 
loading can be considered as if there 
were a plasma which flows upwards from the neutral line. 
In the variant of the model where the magnetic field 
consists in 
closed field lines along a neutral line,
loading can be considered as if there were a plasma which expands
away from the neutral line.

\smallskip
\n {\bf 4.\ Small scale processes (bursts):} 
The redistribution events occurring at unstable sites can be considered 
as localized diffusion processes, accompanied by energy release through 
current-dissipation. The diffusion is accomplished in one-time step, going 
from the initial state directly to the asymptotic solution of a simple 
diffusion equation. The diffusivities, diffusive length-scales and diffusive 
times are the same for all bursts. 

\smallskip
\n {\bf 5.\ Spatio-temporal evolution of the electric field:}
The X-CA model yields the spatio-temporal evolution of the
intense and localized electric fields, which appear at the sites of 
current-dissipation during flares.
Typically, the electric fields are of similar magnitude and 
similar direction, 
and the locations where they appear travel through the 
grid in the course of time.

\smallskip
\n {\bf 6.\ The current as the critical quantity:}
A modification which brings the X-CA model closer to Parker's
flare scenario and plasma physics is the 
replacement of the standard stress measure with the 
current, so that 
directly a large current is responsible for the occurring of a burst. 
First results indicate that the SOC state basically 
persists under this modification, the large scale structure of the magnetic 
field remains the same,
the distributions of total and peak energy remain power-laws, with a 
slight tendency towards steepening. 

\smallskip 
\n {\bf 7.\ The nature of the instability criterion and the diffusivity:}
The local diffusion events start if 
a locally defined stress exceeds a threshold.
This local stress corresponds either to an approximately 
linear function of the 
current for large stresses (in the standard version of the X-CA model; 
see Sec.\ 3.3), 
or directly to the current (in the version of Sec.\ 4).
The X-CA model thus implicitly implements Parker's flare scenario that an 
instability is triggered if the current $\vec J$ (or a linear function of it) 
exceeds some threshold, with the result that 
the resistivity increases and diffusion dominates the time evolution.
Physically, one would think of the diffusion to become anomalous; in the 
X-CA model the resistivity switches locally from zero to one 
during one time-step.

\smallskip
\n {\bf 8.\ Global organization of the current-dissipation regions:} 
The current-dissipation is spatially and temporally fragmented into a large 
number of practically independent, dispersed, and disconnected  
dissipation regions with the 
shape of current-surfaces, which vary in size and 
are spread over a considerable volume. These current-surfaces do not grow
in the course of time,
but they multiply and are short-lived.

\subsection{Discussion}

The magnetic topology in the X-CA model (Sec.\ 3.1.2) has to be 
compared to 
the current picture we have of a flaring active region, where the field 
topology is complex,
with structures on all scales, and with no simple organization of the entire 
flaring region. 
A judgment of the X-CA model's magnetic field topology depends on what part of 
an active region one intends to describe. 
If we assume or intend to model entire active regions or 
substantial parts of them, then we would naturally 
prefer the variant of the X-CA model where the magnetic field forms an
arcade of field lines (Fig.\ 2).
Qualitatively, the picture the model gives is not bad, though the observations
show a still higher degree of complexity (more than one, and non-straight 
neutral lines, etc.). Moreover,
it seems unlikely that well separated, isolated loops can be identified in the
model's magnetic field structure. 
These two discrepancies should preferredly be interpreted as 
simplifications the model makes --- although, they alternatively
might also be interpreted in the way that the magnetic topology 
represents only a part of an active region, or even just the inner part of 
one single loop. However, this second interpretation would just open new 
questions of adequacy, which replace the discussed ones. 

More difficult to judge is what 
the magnetic 
field topology 
of the standard variant of the model, 
the closed field-lines along a straight neutral line (Fig.\ 1),
might correspond to. 
Such structures are not observed, so that they would have 
to correspond to small-scale structures, below today's observational 
capabilities. We might, for instance, assume that these structures are
the X-CA model's representation of an eddy of three-dimensional MHD turbulence.

The variant of the X-CA model which yields the arcade of field lines
has physically more realistic boundary conditions 
(closed boundaries at the bottom plane; Secs.\ 2 and 3.1.2) 
than the standard form (open boundaries at the bottom plane),
if we assume the bottom plane to
represent the photosphere: 
Coronal flares (avalanches)
may propagate out of the simulation cube in all directions, assuming
that we are not modeling the entire corona, they should, however, 
not propagate freely into the photosphere, where the physical
conditions are strongly different from the ones in the corona, 
but they should rather leave the photospheric magnetic field 
basically unchanged.
Note that the discussed boundary conditions are relevant in our model 
(as well as in the classical CA models) only for 
the bursts, not though for loading, which we discuss next.

The loading process has the interesting interpretation that it implicitly
assumes a velocity field which systematically flows upwards against the
arcade of magnetic field-lines (or expands the closed field lines, in the 
case of the other
magnetic topology), 
which is very reminiscent of the realistic scenario of newly
emerging, upwards moving flux, pushing against the already existing
magnetic flux and causing in this way occasional magnetic
diffusion events, i.e.\ events of energy release (Sec.\ 3.2).
Despite this interesting interpretation, 
the loading process is 
still unsatisfyingly simplified:
(a) The loading increments $\delta \vec A$ 
do depend on the direction of the pre-existing magnetic field (see Sec.\ 3.2),
but they should also depend on the magnitude of $\vec B$
if one assumes them to represent disturbances according to the 
$\vec v \wedge \vec B$ term of the induction equation. 
(b) The loading process acts everywhere and independently 
in the entire simulation box, whereas according to Parker's flare scenario
(see App.\ A), 
it should act independently only on one boundary of the simulation box
and propagate from there into the system, since
an active region is driven only from one boundary, the photo-sphere,
(by random foot-point motions and newly emerging flux), from where 
perturbations propagate along the magnetic field-lines into the active region.
We just note that 
also all the more or less different
loading processes of the classical CA models suffer from the 
problems (a) and (b).
A velocity field was explicitly introduced into a CA so far only by the 
CA model of Isliker et al.\ (2000a), which
is, however, a non-classical CA model, with evolution rules
directly derived from MHD.

An interesting property --- or prediction --- of the X-CA 
model is
the preferred directionality of the appearing currents and electric fields,
parallel to the neutral line (Secs.\ 3.1.1, 3.4). 
Since both the currents and the electric 
fields are only indirectly observable, this prediction is difficult 
to verify with observations. 
The length-scale over which the currents and electric fields are parallel
depends on what part of an active region 
the X-CA model actually represents.

It is also worthwhile noting that the currents are everywhere more or less
perpendicular to the magnetic field (Sec.\ 3.1.1), and therewith the magnetic 
field
in the physical set-up of the X-CA model is not force-free, opposite
to what is usually assumed in MHD for the coronal plasma in its
quiet evolution.
As the current, so is the electric field always 
more or less perpendicular to the magnetic field, having in general, though,
a small parallel component (Sec.\ 3.4).

The model's diffusive small-scale physics in the burst
mode represents quite well anomalous diffusive processes, despite 
some characteristic 
simplifications (Sec.\ 3.3). 
The most peculiar assumption made in the X-CA model is the conservation
law for the 
vector-potential ($\int \vec A \,dV =const.$), which holds 
during bursts and 
which is a necessary condition for the X-CA model, as for the
classical CA models, to reach the SOC state 
(see e.g.\ Lu \& Hamilton 1991; Lu et al.\ 1993).  
As a consequence, also $\int \vec B \,dV$ is conserved during bursts.
The physical meaning of this conservation law seems unclear:
in MHD, for instance, not directly $\int\vec A\,dV$ or $\int\vec B\,dV$ are 
expected to be conserved, but $\int\vec A \vec B \, dV$, the magnetic helicity 
(if the integration volume is chosen adequately; see e.g.\ Biskamp 1997).

The regions of intense, but sub-critical current-density in the quiet 
evolution of the X-CA model are organized in current surfaces of various sizes
(Sec.\ 5). 
A similar picture, though with characteristic differences (e.g.\ 
with much less fragmentation), has been reported 
in the 3-D MHD simulations of coronal plasmas by Nordlund and Galsgaard 
(e.g.\ 1997). The
pictures yielded by the X-CA model and by the MHD simulations
are different not least due to the fact that 
the MHD simulations have high spatial resolution, and they model
a smaller volume than the X-CA model does, so that, among others, the current 
surfaces in the X-CA model are spatially less resolved, they are smaller, and 
they do not 
reach the size of the entire simulation box as they do in the MHD simulations.

The current-dissipation regions at any time during 
a flare in the X-CA model 
do not show any sign of global spatial organization between them, and 
they can definitely not be considered 
as the dissipation and destruction of a well defined, simple
structure (as for instance the disruption of a single, extended 
current-sheet would be). Moreover, the energy dissipation shows a highly 
dynamic spatio-temporal behaviour: The current-dissipation regions
are not statically maintained at fixed grid-sites during a flare 
(as it would be the case if they were continuously fed with in-streaming 
plasma), but they are short lasting and 
travel through the grid, exploring the near-to-unstable regions.
As a consequence, the volume participating in the energy release 
process is considerably large at most times during a flare, a flare in the 
X-CA 
model is never a localized process. Lastly, note that all the ever 
changing 
current-dissipation regions which participate in a flare  
carry their own, independent magnetic field-lines, which 
are rooted in the photosphere 
(in the variant of the model with the magnetic field topology 
in the form of an arcade, Fig.\ 2).

Finally, it is worthwhile noting an essential difference 
between MHD simulations and the X-CA model:
MHD simulations do not so far invoke anomalous resistivity.
In MHD simulations, 
$\eta$ is given a fixed and constant numerical value (which
moreover is usually adjusted to the grid-size for numerical reasons).
The X-CA model, on the other hand, incorporates the kinetic plasma
physics which rules the behaviour of the resistivity $\eta$, simulating the 
effect of occasionally appearing anomalous resistivities due to current 
instabilities (see Sec.\ 3.3).
As all the classical CA models, it can so far not model 
current dissipation in the frame of a constant, ordinary diffusivity 
as the result of the interplay of shears in the magnetic field
and the velocity field. A complete model for solar flares should
ultimately incorporate both dissipation mechanisms.

Due to this difference, 
a comparison of the current-dissipation regions of the X-CA model in the 
flaring
phase to MHD simulations seems not realistic.

\subsection{Conclusions}

The X-CA model represents an implementation of Parker's (1993)
flare scenario, covering aspects from 
small-scale plasma physics and MHD to the large scale physical set-up and 
magnetic topologies:
most aspects 
are in good
accordance with Parker's flare scenario, even though some give rise to 
ambiguous interpretations with associated open questions, and some 
involve unsatisfying simplifications which need improvement. 
One should be aware  
that CA models, which by definition evolve according to rules in a discrete
space and in discrete time-steps, have  
by their nature to make simplifications,
and one cannot expect them to give exactly the same picture as the 
observations or MHD simulations, one can just demand that the 
simplifications are adequate and reasonable, that the over-all picture 
is as close as possible to the physical one, and, of course, 
that the quantitative results they give (e.g.\ concerning energy release)
are in good accordance with the observations.

The X-CA model 
allows different 
future applications and questions which could not be asked so far in the frame 
of classical CA models, and it gives more or refined results. 
One application is a more detailed comparison of the X-CA model
to observations. For instance, particles can now be introduced into the
model, their thermal radiation can be monitored,
and they can be accelerated 
through the electric fields to yield non-thermal emission 
(e.g.\ synchrotron emission; an earlier
study of particle acceleration in a classical CA model was made by 
Anastasiadis et 
al.\ (1997), who had to estimate the electric field still indirectly).
Very promising on the side of the X-CA model is that the energy 
dissipation is fragmented and
spread over a considerably large volume, with a large number of 
dissipation regions, so that particle acceleration in the frame
of the X-CA model can be expected to be very efficient.

An important property of the X-CA model is not least its flexibility,
which allows to implement 
concrete plasma-physical or MHD ideas in the frame-work of a CA.
This was demonstrated here and in IAV2000 by several modifications:
the direct use of the current in the instability criterion, 
the energy release in terms of Ohmic dissipation, and
by the modifications which led to a more realistic magnetic topology.

\begin{acknowledgements}
We thank K.\ Tsiganis and M.\ Georgoulis for many helpful discussions on 
several issues. We also thank G.\ Einaudi for stimulating discussions on
MHD aspects of flares,
and the referee A.L.\ MacKinnon for discussions on several aspects of CA
and MHD, and for his critics which helped to improve this article.
The work of H.\ Isliker was partly supported by a grant of the Swiss National 
Science Foundation (NF grant nr.\ 8220-046504).
\end{acknowledgements}

\appendix

\section{Short summary of Parker's flare scenario}

The flare scenario of Parker (e.g.\ 1993) can briefly be summarized as follows
(whereby also a few basic observational facts concerning flares and
active regions shall be mentioned):
Active regions are characterized by a highly complex magnetic topology, with 
sub-structures on a large variety of scales 
(e.g.\ Bastian \& Vlahos 1997, Bastian et al.\ 1998). 
Generally, in an active region the diffusivity $\eta$ is small (the magnetic 
Reynolds number is much larger than unity), and convection 
dominates the evolution of the magnetic field, i.e.\ 
the magnetic field is built-up and continuously 
shuffled due to random photospheric foot-point 
motions (the magnetic fields are ultimately rooted in the turbulent 
convection zone). In this way, magnetic energy is stored in active
regions. Occasionally, magnetic structures with high shear may 
locally be formed, in which the current is increased. If the current is 
intense enough, then it is expected from plasma-physics that a kinetic 
instability is triggered, most prominently the ion-acoustic instability. This 
instability causes in turn the diffusivity $\eta$ of the plasma to become 
locally anomalous and therewith to increase 
drastically (by several orders of magnitude, see e.g.\ references in Parker 
(1993)). The evolution of the magnetic field is then 
governed {\it locally} by diffusion, convection is negligible.
In these local diffusion processes, energy is released due to Ohmic 
dissipation with a rate $\eta \vec J^2$, until the free energy is more or 
less dissipated and the current has fallen to a much smaller value, so that
also $\eta$ returns to its ordinary value.  
In flares, such local diffusion events (bursts) appear 
in a large number during a relatively short 
period of time, spread over this time-interval and in space, and releasing 
in their sum considerable amounts of energy. Flares are thus considered to be
fragmented into many sub-events, and there is some kind of chain-reaction
or domino-effect, whose exact form is an open problem of flare modeling 
(CA models for instance consider a domino-effect to be operating).

\section{Open and closed boundary conditions}

The boundary conditions (b.c.) around the simulation cube affect
the redistribution rules and the definition of the stress measure 
$\vec S_{ijk}$. In case of open b.c., an implicit layer of zero-field around 
the grid is assumed, held constant during the entire time-evolution. 
The numerical factor
$n_n$ in the definition of $d \vec A_{ijk}$ and in 
the redistribution rules (see Sec.\ 2) has a fixed value, $n_n=6$, 
assuming that every grid point has six nearest neighbours (the grid
we use is cubic),
independent of whether it is at the boundaries or not. 
Consequently, in the definition of $d\vec A_{ijk}$  
the sum has always six terms, 
the $\vec A_{nn}$ outside the grid contributing zero. 
The 
continuation method which is used to determine $\vec B$ and $\vec J$
explicitly takes the zero-layer around the grid into account
(see IAV2000).

In the case of closed b.c., no communication takes place between 
the field in the grid and the region outside the grid.
The definition of $d \vec A$ is adjusted to $d\vec A = 
\vec A - 1/m_n \sum^\prime \vec A_{nn}$,
where the primed sum is now only over the nearest 
neighbours which are {\it inside} the grid, and $m_n$ is the number of 
these interior nearest neighbours 
($m_n$ can thus be less than $6$). 
The continuation method does not assume 
any layer of pre-fixed field around the grid in order to determine
$\vec B$ and $\vec J$.
The redistribution rules are formally the same as introduced
in Sec.\ 2, just that again $n_n$ is replaced by $m_n$, the effective
number of nearest neighbours inside the grid.

As stated in Sec.\ 2, we use two version of b.c., one where all the
boundaries are open, and a mixed b.c., with open boundaries at all the 
boundary planes except for a closed boundary at the lower $x$-$y$ plane,
i.e.\ we assume a layer of zero-field around the grid and take it into
account, except at the lower boundary, which is treated differently,
as described above.

\end{document}